\documentclass[submitting]{nst}
\usepackage{subfigure,dcolumn}
\usepackage[T2A,T1]{fontenc}
\usepackage[russian,english]{babel}
\usepackage{verbatim}
\usepackage{makecell}
\usepackage{hyperref}
\usepackage{multirow}
\usepackage{unicode-math}
\usepackage{listings}

\usepackage{ulem}
\usepackage{etoolbox}
\robustify\uline

\makeatletter

\newcommand{\Rmnum}[1]{\expandafter\@slowromancap\romannumeral #1@}
\def\@fnsymbol#1{\ensuremath{\ifcase#1\or \dagger\or \ast\or \ddagger\or \mathsection\or \mathparagraph\or \|\or **\or \dagger\dagger\or \ddagger\ddagger \else\@ctrerr\fi}}
\makeatother

\usepackage{lineno} 
\makeatletter
\let\oldmaketitle\maketitle
\renewcommand{\maketitle}{%
  \oldmaketitle
  \nolinenumbers
}
\makeatother

\begin{document}

\title{Hybrid Electromagnetic Calorimeter Module: Enhanced Performance through Integration of Silicon Pixel Layer into Scintillator Design}

\author{Jia-Le Fei}\thanks{These authors contributed equally to this work}
\affiliation{The Institute for Advanced Studies, Wuhan University, Wuhan 430072, China}
\affiliation{College of Naval Architecture and Ocean Engineering, Naval University of Engineering, Wuhan 430033, China}

\author{Ao Yuan}\thanks{These authors contributed equally to this work}
\affiliation{The Institute for Advanced Studies, Wuhan University, Wuhan 430072, China}

\author{Chang-Heng Huang}\thanks{These authors contributed equally to this work}
\affiliation{The Institute for Advanced Studies, Wuhan University, Wuhan 430072, China}

\author{Liu-Pan An}
\email[Corresponding author, ]{E-mail address: anlp@pku.edu.cn}
\affiliation{School of Physics State Key Laboratory of Nuclear Physics and Technology, Peking University, Beijing 100871,
China}
\author{Ji-Ke Wang}
\email[Corresponding author, ]{E-mail address: jike.wang@cern.ch}
\affiliation{The Institute for Advanced Studies, Wuhan University, Wuhan 430072, China}

\begin{abstract}
This study constructs a hybrid module with a multi-material collaborative detection architecture by integrating silicon pixel layers into the longitudinally segmented scintillating fiber sampling calorimeter module and optimizes the placement of the silicon layers. The module utilizes the pre-shower characteristics of the front-end scintillator units to ensure sufficient energy deposition in the silicon pixel layers, thereby maintaining its high-precision detection capability. A dedicated simulation framework combining Geant4 modeling for the scintillator section and a parameterized approach for the silicon pixel layer is employed for module verification and performance study. The hybrid module demonstrates overall performance enhancement. The maximum achievable improvements are 56\% for position resolution and 26\% for time resolution, respectively. These advancements also lead to significant increase of physics sensitivity, especially for physics channels with low-energy photons, for instance, a 16\% boost in signal significance for $D^{*0}$ from the $B^-\rightarrow D^{*0}(\rightarrow D^0 \gamma)\pi^-$ decay.

\end{abstract}


\maketitle

\section{Introduction}
\begin{figure}[!htb]
    \centering
    \includegraphics[width = 8.5cm]{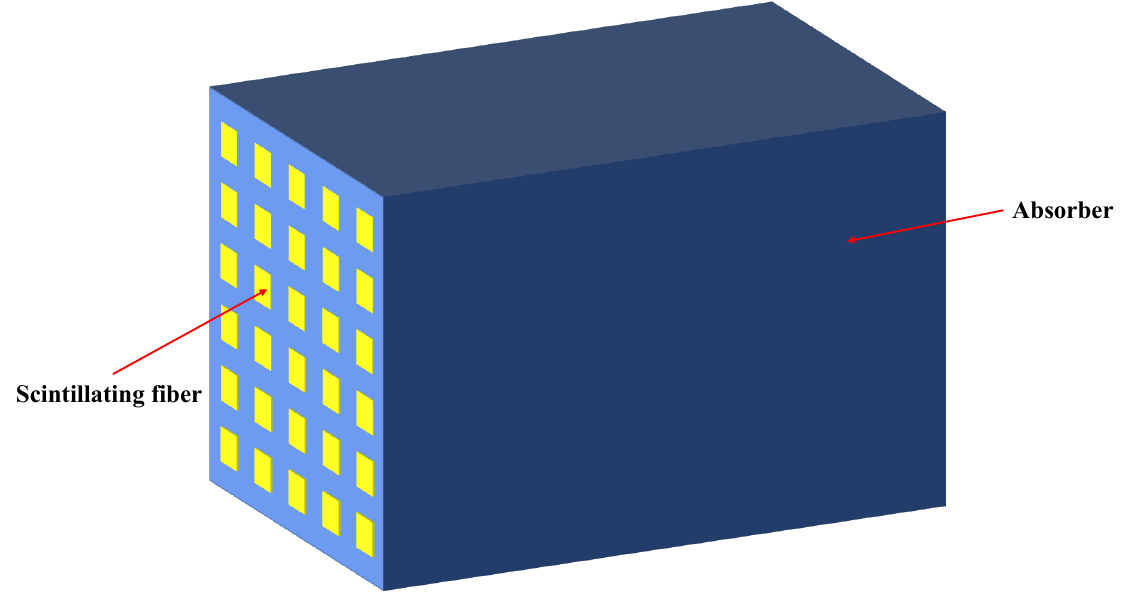}
    \caption{Diagram of the SpaCal module.}
    \label{fig:SPACAL_Module}
\end{figure}

Current research in particle physics, supported by a growing number of accelerators worldwide, is advancing along two main directions: the energy frontier and the luminosity frontier. Among these, high-luminosity collider experiments are gradually emerging as the central focus of research, owing to their potential to provide unprecedented precision in exploring fundamental physics. High-luminosity collider experiments present significant technical challenges to detector systems due to the extreme particle flux densities involved\cite{Future_Experiment}. For instance, the upcoming High-Luminosity Large Hadron Collider (HL-LHC) will increase the instantaneous luminosity by a factor of five compared to the present LHC operation, creating an experimental environment with extremely high occupancy and background rates\cite{HL_LHC}. Similar challenges are expected in Belle II at KEK, which targets an integrated luminosity of 50 $ab^{-1}$\cite{BelleII_PhysicsBook}, and in proposals such as the Future Circular Collider (FCC-ee)\cite{FCCee_Overview}. These projects impose stringent requirements on calorimeter modules in terms of granularity, time resolution, and radiation hardness. As a critical component of the particle detection systems, electromagnetic calorimeters (ECAL) specialize in measurement of photons and electrons\cite{ECAL_1,ECAL_2}, while enabling indirect reconstruction of neutral mesons like $\pi^0$ through secondary decay chains. There are two principal architectures of ECAL: total absorption and sampling calorimeters. In high-luminosity collider experiments, sampling calorimeters dominate due to their small Molière radius, which ensures good energy resolution and effective $\pi^0-\gamma$ separation, typically employing high-Z metals (e.g., lead and tungsten) as absorbers\cite{Sampling_Cal,MOLIERE}. Sampling calorimeters typically adopt a sandwich-type architecture with alternating stacks of absorber layers and active layers. Such a design allows the absorber to induce particle showers while the active medium samples the deposited energy, thereby achieving a balance between performance and cost. When scintillating crystals are employed as the active layer, this configuration is specifically termed the Shashlik structure, as it utilizes longitudinally embedded optical fibers to channel scintillation photons to the readout electronics. Collider experiments in high-luminosity environments are imposing stringent new performance criteria on calorimeter modules, specifically requiring enhanced radiation resistance, more compact design with smaller Molière radius, and greater flexibility in readout cell dimensions. Conventional Shashlik sampling structures struggle to meet the new experimental requirements in the core region adjacent to the beam pipe. Therefore, some new sampling structures are being adopted. As illustrated in Fig.~\ref{fig:SPACAL_Module}, the proposed design integrates scintillating fibers into perforated absorber plates, allowing the fibers to serve dual functions as both active sensing elements and optical transmission channels. This configuration achieves concurrent optimization of three critical parameters: (1) increased absorber mass fraction for Molière radius ($R_M$) and radiation length ($X_0$) reduction; (2) diminished light scattering and light attenuation; (3) enhanced radiation resistance through the geometric advantages of fiber-shaped scintillators. Owing to the application of scintillating fiber technology, this geometric configuration is formally designated as “SpaCal” (Spaghetti Calorimeter)\cite{SPACAL_Radiation,SPACAL_prototype,SPACAL0,SPACAL1,SPACAL2,SPACAL3,SPACAL4,SPACAL5}. SpaCal technology has already been adopted in several test-beam campaigns and detector upgrades, including prototypes for the LHCb ECAL Upgrade II \cite{PicoCal_JINST} and the forward calorimeter of the STAR experiment \cite{STAR_FCS}. These practical examples demonstrate its feasibility, yet they also reveal intrinsic limitations in terms of spatial granularity and uniformity. In particular, the achievable detector cell size is constrained by photon number fluctuations during light transport and by the physical dimensions of photo-conversion devices such as photomultiplier tubes or SiPM arrays.

To overcome these limitations and achieve even finer spatial granularity, alternative technologies are being explored. Silicon-based semiconductor detectors exhibit fast response time and low carrier fluctuation, offering better linearity in response and precise timing measurement\cite{time_pos_1,time_pos_2,time_pos_3}. Meanwhile, the direct charge collection mechanism, on-chip readout technology, and mature semiconductor fabrication processes of the silicon-based semiconductor detector enable it to achieve a smaller unit size both in transverse and longitudinal dimension\cite{technology_1,technology_2,technology_3}, while still maintaining excellent performance. However, if they are used to cover sufficient radiation length, their fabrication costs remain significantly higher compared to scintillator-based systems.

Notably, the advantages of silicon pixel and strip detectors have been demonstrated by a wide body of research\cite{Silicon_charge_det,Silicon_heavy_ion_exp,Silicon_3,Silicon_4,Silicon_5,Silicon_6}.~The CMS High Granularity Calorimeter (HGCAL)\cite{CMS_HGCAL_TDR} employs silicon sensors in its endcap region to achieve unprecedented spatial resolution, while the ALICE Inner Tracking System (ITS)\cite{ALICE_ITS_TDR} demonstrates the scalability of pixel technology in harsh environments.~ATLAS has also developed the High-Granularity Timing Detector (HGTD)\cite{ATLAS_HGTD_TDR} to cope with pile-up by providing precise time information. These examples illustrate the potential of silicon technology, but also highlight its prohibitive cost when considered as a standalone absorber system.

By integrating a multi-material collaborative detection architecture, the respective advantageous detection characteristics of different materials can be effectively combined to enhance the overall performance of the module. This study proposes a hybrid electromagnetic calorimeter module design that combines SpaCal-type scintillator units with silicon pixel detectors in an integrated architecture. The design strategically leverages silicon’s inherent advantages in fast timing response and fine spatial granularity while maintaining scintillator-based energy measurement capabilities. A simulation framework integrating Monte Carlo methods with parametrization techniques is established for the hybrid module. Comprehensive validation is conducted across a single-photon bench and multiple physics decay channels, demonstrating baseline performance.

\section{Design of hybrid Module}

In early-generation, ECAL was primarily designed to capture transverse shower profile and energy deposition, while lacking sufficient granularity to resolve detailed longitudinal shower development\cite{LEP1,LEP2,LHCb_ECAL}. The precise reconstruction of the longitudinal development characteristics of electromagnetic shower plays a critical role in particle identification and time measurement accuracy, necessitating a longitudinal multi-layer sampling structure in calorimeters. Consequently, longitudinal layered sampling technology has become the mainstream solution. The CMS’s HGCAL endcap calorimeter, ALICE’s FoCal detector, and LHCb’s upgraded PicoCal all adopting this paradigm\cite{HGTD_ATLAS,CMS_ECAL_PID,HGCAL,FoCal,LHCbU2_ScopingDocu}.

Through longitudinal segmentation and dual-end readout design, the SpaCal module can add an additional layer of granularity along the longitudinal direction. The segmentation interface precisely provides the necessary space for installation of silicon pixel layers.

The combination of absorber and active materials in the SpaCal module significantly impacts its performance. This study introduces silicon pixel layers into the SpaCal module, with particular emphasis on analyzing their effects on the performance of SpaCal modules with different material systems. Two representative combinations are selected as research subjects: the SpaCal modules constructed with W-GAGG and Pb-Polystyrene (absorber-active material) configurations. Compared to the Pb-Polystyrene configuration, the W-GAGG configuration exhibits a smaller Molière radius\cite{SPACAL_prototype,SPACAL0}.

\begin{figure}[!htb]
    \centering
    \includegraphics[width = 8.5cm]{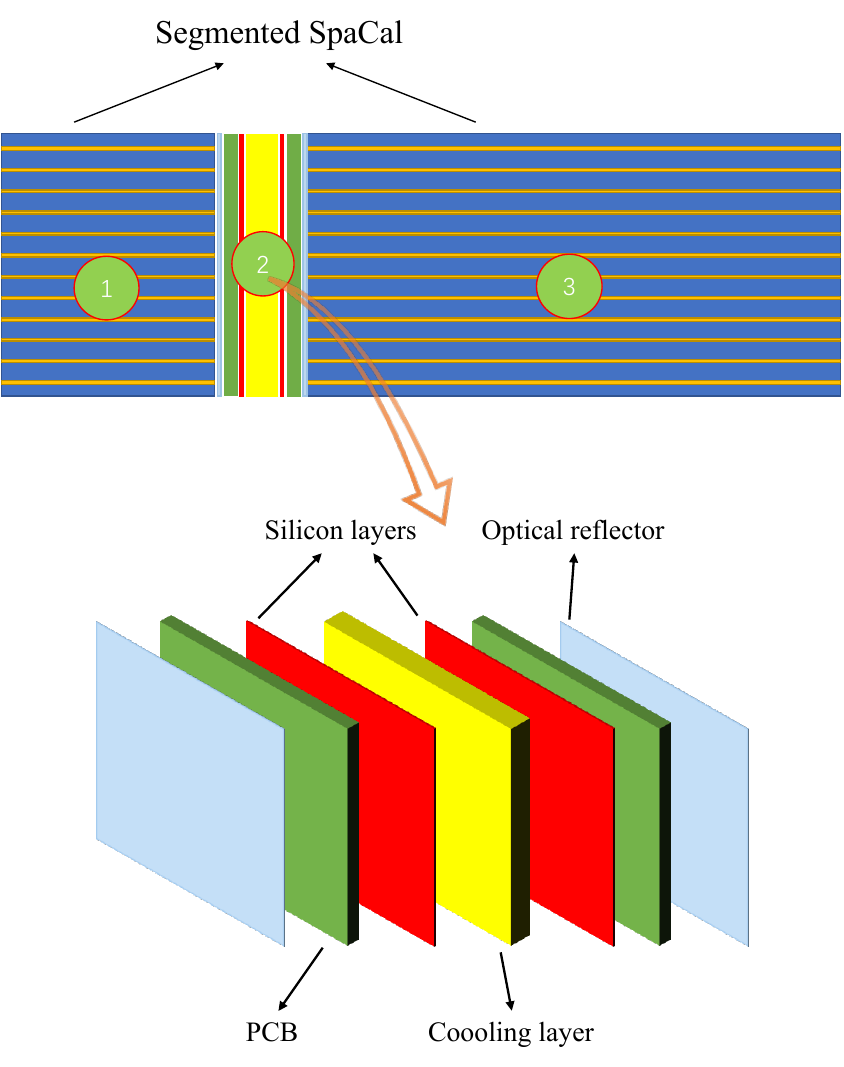}
    \caption{The illustration of the SpaCal-Silicon module.}
    \label{fig:SPACAL-Silicon}
\end{figure}

As conceptualized in Fig.~\ref{fig:SPACAL-Silicon}, the silicon pixel region (located at position 2) is composed of two pixel layers (red) at the SpaCal segmentation boundary, sharing a unified copper cooling layer (yellow). Substrate PCBs manufactured from FR-4 material flank both sides of the pixel layers (green). Optical reflective films (cyan) are also used adjacent to the SpaCal Zones (zones 1 and 3) to optimize light collection efficiency. In the following text, we refer to this hybrid module as the SpaCal-Silicon module.

The design philosophy of this hybrid architecture is based on the principle of material complementarity. While scintillating fibers embedded in absorber plates provide a cost-effective solution for large-volume energy measurement with good light yield, silicon pixel layers contribute fine granularity, low electronic noise, and precise timing. By embedding silicon layers exactly at the shower development stages where the density of minimum ionizing particles (MIP) is highest, the calorimeter gains access to detailed spatial and temporal shower information without sacrificing the total radiation length required for energy containment. This "division of labor" between materials allows the module to combine the strengths of both technologies while minimizing their individual drawbacks.

\begin{figure}[!htb]
    \centering
    \includegraphics[width = 8.5cm]{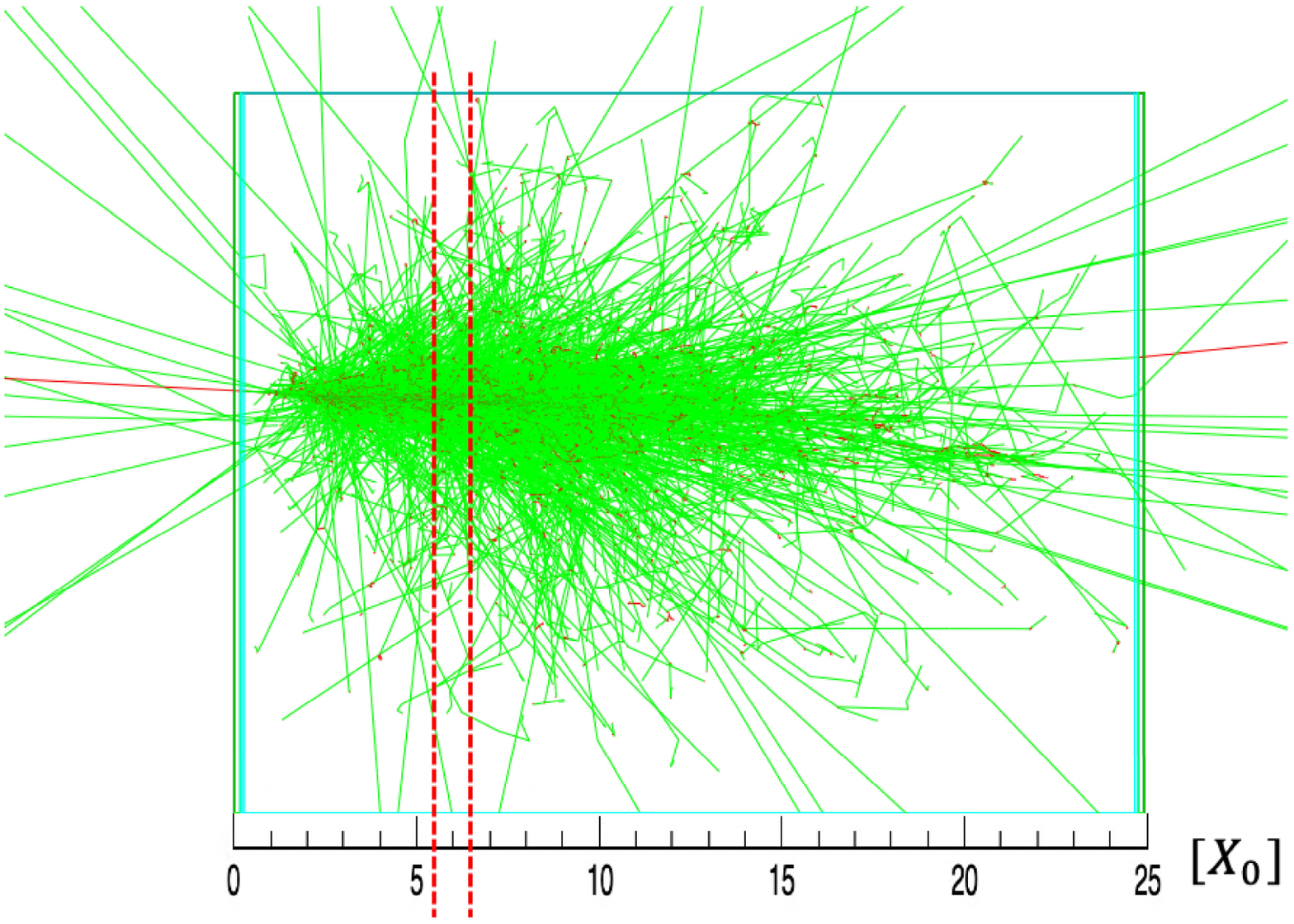}
    \caption{When a particle enters the SpaCal module, it initiates electromagnetic shower development.}
    \label{fig:SPACAL Showering}
\end{figure}

Fig.~\ref{fig:SPACAL Showering} illustrates the development of electromagnetic showers in SpaCal modules. The performance of silicon pixel layers depends on the density of minimum ionizing particle (MIP), which is governed by the longitudinal development stage of electromagnetic showers. Meanwhile, the transverse size of the shower is also closely related to the longitudinal development stage of the electromagnetic shower. Consequently, their longitudinal placement must be balanced between the following two considerations and is inherently related to the position where shower maximum ($X_{\text{max}}$) is.
\begin{itemize}
    \item \textbf{Electron density}: Increases with proximity to shower maximum (higher particle flux near $X_{\text{max}}$).
    \item \textbf{Shower splitting capability}: Enhances with distance from $X_{\text{max}}$.
\end{itemize}

\begin{figure}[!htb]
    \centering
    \includegraphics[width = 8.5cm]{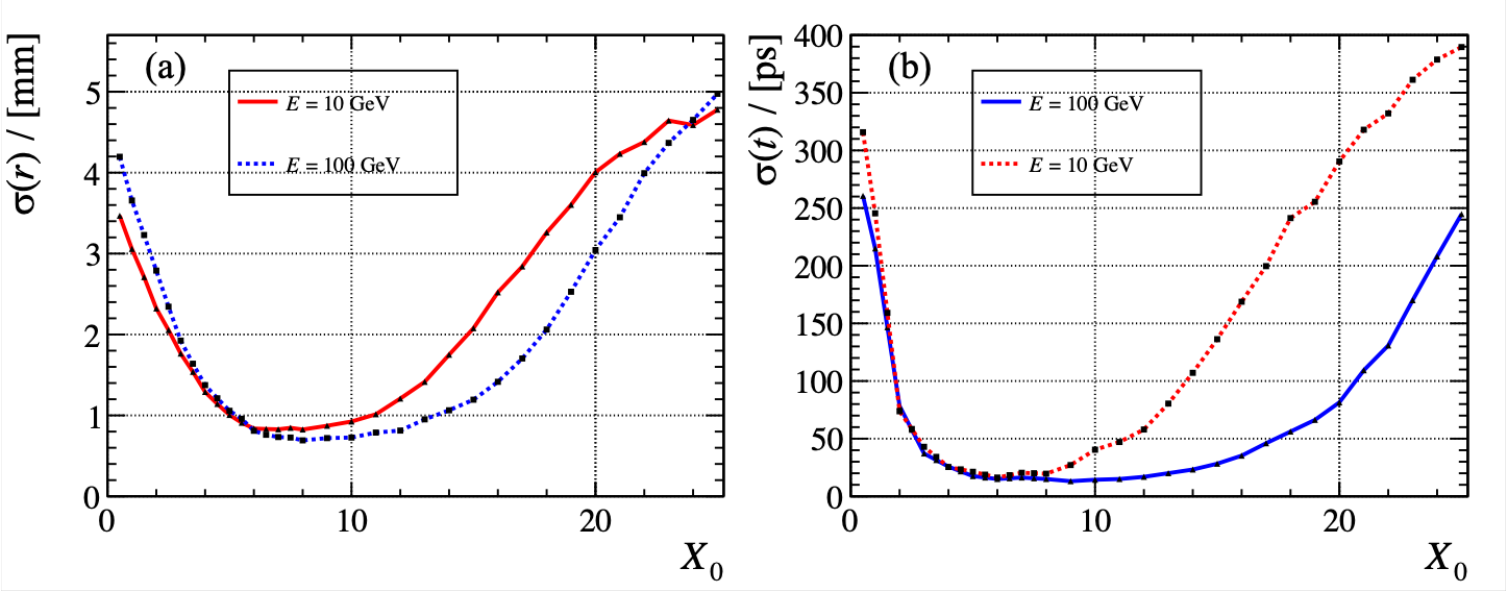}
    \caption{Functional dependence of single-layer resolution on longitudinal positioning parameters in the silicon layer (expressed in radiation length units $X_0$). (a): position resolution. (b): Time reoslution.}
    \label{fig:Silicon_Resolution}
\end{figure}

To optimize the position configuration of the silicon layer, this study dynamically adjusts its longitudinal positioning by changing the thickness of the front-end SpaCal structure. As demonstrated in Fig.~\ref{fig:Silicon_Resolution}, the diagram systematically reveals the quantitative correlation between the longitudinal position and the time (position) resolution of the silicon layer. Fig.~\ref{fig:Silicon_Resolution} demonstrates that the time (position) resolution achieves its optimal operational window within about 5-8 $X_0$ radiation lengths. Therefore, positioning the silicon pixel layers within the 5-8 $X_0$ range is identified as the optimal configuration (marked by red dashed lines in Fig.~\ref{fig:SPACAL Showering}).

Based on the aforementioned two SpaCal module configurations (W-GAGG and Pb-Polystyrene), we construct two types of SpaCal-Silicon hybrid modules. The 5 mm pixel pitch selection achieves smaller cell size than Molière radius of the module while maintaining manageable readout channel counts. The key module parameters are summarized in Table~\ref{tab:SPACAL-Silicon模块}.

\begin{table*}[!htb]
    \centering
    \caption{SpaCal-Silicon Modules}
    \label{tab:SPACAL-Silicon模块}
    \footnotesize
    \begin{tabular*}{14cm} {@{\extracolsep{\fill} } lcc}
        \toprule
        SpaCal-Silicon Modules & W-GAGG-Si & Pb-Polystyrene-Si \\ \hline
        Absorber material & Tungsten & Lead \\ \hline
        Crystal material & GAGG & Polystyrene \\ \hline
        Thickness of Silicon layer (mm) & 0.5 & 0.5 \\ \hline
        Thickness of Cooling layer (mm) & 6 & 6 \\ \hline
        Thickness of PCB (mm) & 3.6 & 3.6 \\ \hline
        Size of pixel cell ($\mathrm{mm}^2$) & $5 \times 5$ & $5 \times 5$ \\ \hline
        Size of SpaCal cell  ($\mathrm{mm}^2$) & $15 \times 15$ & $15 \times 15$ \\ \hline
        Thickness of Front SpaCal (mm) & 35 & 75 \\ \hline
        Thickness of Back SpaCal (mm) & 105 & 210 \\ \hline
        Number of layers & 4 & 4 \\
        \bottomrule
    \end{tabular*}
\end{table*}

\section{Simulation}
During the development of calorimeter modules, simulation plays a crucial role in validating the feasibility of the design, optimizing performance, and reducing both cost and technical risk. The simulation of the SpaCal-Silicon module consists of two parts: the SpaCal module and the silicon pixel layer. For the SpaCal module, a simulation framework\cite{SimCode}developed by the LHCb ECAL Upgrade group based on Geant4\cite{Geant4}is employed to model particle interactions, scintillation light generation, and light transport within the fiber-embedded absorber plates. Additional software development is required for the silicon pixel layer, which involves simulating the complex charge collection and signal formation processes.

\begin{figure}[!htb]
    \centering
    \includegraphics[width = 8.5cm]{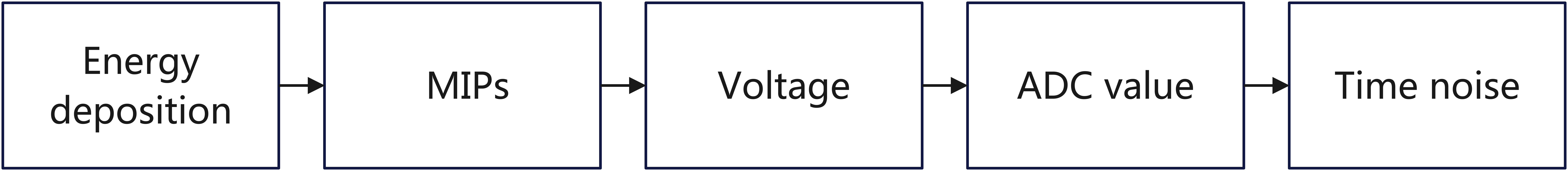}
    \caption{The simualtion flow chart of silicon pixel layer.}
    \label{fig:Silicon层仿真}
\end{figure}

Accurate hardware-level simulation relies heavily on real experimental data to ensure that the models reflect actual detector behavior. Research on silicon detectors\cite{Si_Det} and the development of experimental apparatuses such as HGCAL\cite{HGCAL} and FoCal\cite{FoCal} provide valuable references for benchmarking simulation frameworks. The silicon pixel detectors operate through the drift of ionized electron-hole pairs under an applied bias voltage. However, directly simulating the drift of charge carrier and signal formation is highly complex, as it involves material properties like carrier mobility andconductivity, doping types and methods, bias voltage, and is strongly dependent on the structural design of the semiconductor unit. This paper designs a simplified simulation model based on MIP counts. The simulation workflow is shown in Fig.~\ref{fig:Silicon层仿真}. This section details the modeling approach for the silicon pixel layer.

Energy deposition is simulated using Geant4, requiring the recording of two key parameters from the active layer of each $5\times5\,\mathrm{mm^2}$ readout cell:
\begin{itemize}
    \item Deposited energy within the active volume
    \item Ideal timestamp of the readout unit
\end{itemize}
The implementation follows this formalism:

\begin{equation}\label{eq:能量沉积与时间戳}
    \begin{aligned}
        E_{dep} &= \sum_{\substack{i \\ (x_i,y_i,z_i) \in \Omega}} e_i \\
        t_{tru} &= \frac{\sum_{\substack{i \\ (x_i,y_i,z_i) \in \Omega}} e_i \cdot t_i}{\sum_{\substack{i \\ (x_i,y_i,z_i) \in \Omega}} e_i},
    \end{aligned}
\end{equation}

\begin{figure*}[!htb]
    \centering
    \includegraphics[width = 16cm]{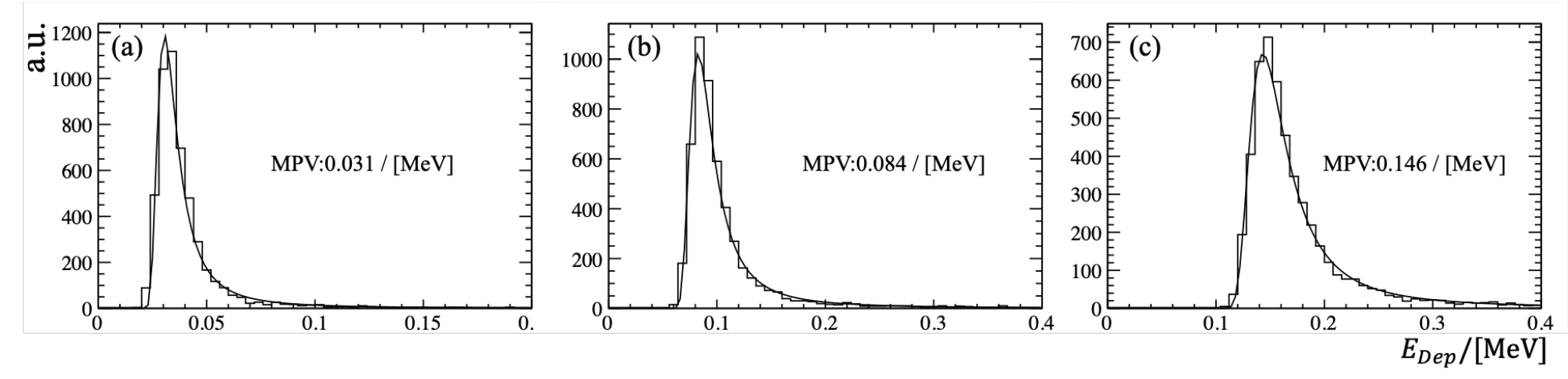}
    \caption{Energy deposition distributions of a MIP in silicon pixel layers with different thicknesses. The MVP represent the most probable value of the Landau distribution. (a): 0.12 mm. (b): 0.3 mm. (c): 0.5 mm.}
    \label{fig:MIPs_Dep}
\end{figure*}

Here, $r_i$ denotes the coordinates of a Geant4 step\cite{Geant4} (where $r$ can be $x$, $y$, or $z$), $\Omega$ represents the spatial extent of the silicon pixel unit's active layer, $e_i$ is the energy of the step, $E_{dep}$ the total deposited energy, and $t$ the ideal timestamp. The number of MIPs traversing the readout unit is then calculated based on $E_{dep}$. The energy deposited by each MIP in silicon follows a Landau distribution. Therefore, we use the most probable value of the Landau distribution as the deposited energy per MIP to estimate the MIP count from the total deposited energy. This first requires determining the energy deposited by a single MIP traversing the silicon pixel unit. While high-energy electrons can be considered MIPs, they generate showers producing numerous secondary particles in material, complicating the measurement of individual MIP energy deposition. Fig.~\ref{fig:MIPs_Dep} shows the energy deposition of MIPs in silicon pixel units.

\subsection{Signal and timestamp generation}
As shown in Fig.~\ref{fig:Silicon层仿真}, after obtaining the number of MIPs passing through each readout unit, we need to further calculate the corresponding output voltage values and simulate the operational process of the analog-to-digital converter (ADC) to complete the analog-to-digital conversion. Finally, the noise in the timing information will be quantified based on the output values of the ADC. Therefore, it is necessary to determine the conversion relationship between the number of MIP and output voltage, and that between the ADC values and time noise\cite{ADC_1}. Additionally, the simulation parameters of the ADC, particularly the sampling depth, reference voltage, and voltage noise, need to be determined. In this paper, we make the following configurations and assumptions regarding the generation of the readout signal and timestamp: 
\begin{itemize}
    \item The output voltage ($V_{out}$) is in a linear relationship with the number of MIPs and the thickness of the active region of the silicon layer. According to Equation~\ref{eq:Out_Voltage}, the output voltage value can be calculated based on the number of MIPs.
    \begin{equation}\label{eq:Out_Voltage}
        \begin{aligned}
            V_{out}=P_1\times N_\mathrm{MIP},
        \end{aligned}
    \end{equation}
    where $P_1$ is a constant parameter, $N_\mathrm{MIP}$ is the number of MIPs.
    \item Here, the timing integration process of the ADC signal is ignored, and we only simulate the digitalization of the output analog voltage from the previous step to obtain the ADC value($\mathrm{ADC}_{out}$). This process can be expressed by:
    \begin{equation}\label{eq:ADC}
        \begin{aligned}
            \mathrm{ADC}_{out}=\mathcal{N}(\frac{V_{out}}{V_{ref}}\times 2^{N_{b}},\sigma_{V}^2),
        \end{aligned}
    \end{equation}
    where, the notation $\mathcal{N}([0],[1])$ represents gaussian sampling with mean $[0]$ and variance $[1]$, $V_{ref}$ represents the reference voltage of the ADC, and $\sigma_{V}$ represents the noise of the reference voltage and $N_{b}$ represents the sampling depth of the ADC. In this paper, we assume that the silicon pixel layer is adequately cooled; therefore, thermal noise affecting the output signal strength is not considered. In addition, shot noise and other noises caused by material defects are also not addressed in this paper.
    \item The noise of time information correlates with the ADC value\cite{Silicon_sensors1}, defined by the following formula:
    \begin{equation}\label{eq:Silicon_noise}
        \begin{aligned}
            \sigma_t=\frac{A}{\mathrm{ADC}_{out}}\oplus C,
        \end{aligned}
    \end{equation}
    where $A$ and $C$ are the parameters related to noise and $\oplus$ is a sum in quadrature. The readout timestamp can be calculated by:
    \begin{equation}\label{eq:Out_Time}
        \begin{aligned}
            t_{out}=\mathcal{N}(t_{tru},\sigma_{t}^2),
        \end{aligned}
    \end{equation}
    where, $t_{tru}$ is the true timestamp obtained using Equation~\ref{eq:能量沉积与时间戳}.
    \item The parameter $A$ in Equation~\ref{eq:Silicon_noise} is in a linear relationship with the thickness of the silicon layer.
\end{itemize}

Based on Ref.~\cite{testBeam_CMS_Silicon,Silicon_sensors1,Silicon_sensors2,Si_Det,Silicon_pixel1,Silicon_pixel2}, for the $5\times5\,\mathrm{mm^2}$ silicon pixel cell with a thickness 0.5 mm under 600 V bias voltage, the parameters of the above configurations and assumptions are listed in Table~\ref{tab:Parameters}.
\begin{table}[!htb]
\centering
\caption{Parameters in simulation}
\label{tab:Parameters}
\footnotesize
\begin{tabular*}{6cm} {@{\extracolsep{\fill} } cc}
\toprule
Parameters & Value \\
\midrule
$P_1$ & 10~mV \\
$N_b$ & 12~bit \\
$V_{\text{ref}}$ & 1~V \\
$V_{\text{noise}}$ & 4.15~mV \\
$A$ & 200~ps$\times \mathrm{ADC}$ \\
$C$ & 14~ps \\
\bottomrule
\end{tabular*}
\end{table}

\subsection{Calibration of the silicon pixel cell}
\begin{figure}[!htb]
    \centering
    \includegraphics[width = 8cm]{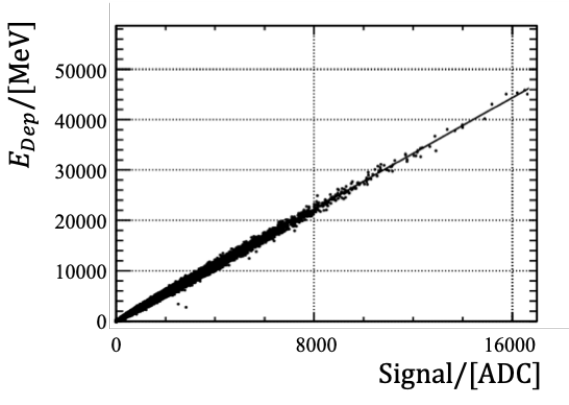}
    \caption{Correlation between silicon pixel layer readout signals and deposited energy.}
    \label{fig:Silicon刻度}
\end{figure}

The calibration of readout units aims to establish the relationship between the readout signal magnitude and the deposited energy within the unit. In the newly integrated silicon pixel units (including silicon pixel readout layer, PCB substrate, and cooling layer), most energy deposits occur in the cooling layer rather than the thin silicon layers. We therefore combine signals from both silicon pixel layers as the total signal, with the energy deposited in silicon pixel units serving as the total energy. As expected, the linear dependence of the output signal on the deposited energy is explicitly illustrated in Fig.~\ref{fig:Silicon刻度}. All silicon pixel units across layers will be calibrated based on this relationship.

\section{Performance}

The primary objective of integrating silicon pixel layers is to enhance both longitudinal and transverse granularity in electromagnetic calorimeter modules. The insertion of silicon pixel layers within the module inevitably diverts a portion of energy originally deposited in SpaCal modules, thereby affecting the performance of downstream SpaCal readout units. Simultaneously, optimizing the placement of silicon pixel layers requires reducing the thickness of upstream SpaCal readout units, which also impacts their performance. The fundamental requirement for silicon layer integration is to improve granularity while minimizing the negative impact on the resolution (especially the energy resolution) of the original module.

Following the layered reconstruction framework in Ref.\cite{Rec}, this section compares the resolution performance between SpaCal-Silicon modules and standard SpaCal modules. The parameters of the standard SpaCal modules are in Table~\ref{tab:Standard SpaCal Modules}. In comparative tests, using the PicoCal in LHCb Phase \Rmnum{2} Upgrade\cite{LHCbU2_ScopingDocu,LHCbU2_FTDR} as a concrete example, the SpaCal-Silicon modules are installed in the layout of PicoCal and completely replace the baseline SpaCal modules (i.e., substituting W-GAGG modules with W-GAGG-Si hybrid modules). Throughout this analysis, "W-GAGG" and "Pb-Polystyrene" specifically denote the standard SpaCal modules.

\begin{table*}[!htb]
    \centering
    \caption{Standard SpaCal Modules}
    \label{tab:Standard SpaCal Modules}
    \begin{tabular*}{14cm} {@{\extracolsep{\fill} } cccccc}
    \toprule
    \thead{Name} & \thead{Type} & \thead{Absorber/active element} &\thead{Cell Size \\ $\left[\mathrm{cm}^2\right]$} & \thead{$R_M$ \\$\left[\mathrm{mm}\right]$} & \thead{Layers}  \\
    \midrule
    W-GAGG & SpaCal & Tungsten/GAGG &1.5$\times$1.5 &14.5 & 2 \\
    Pb-Polystyrene   & SpaCal & Lead/Polystyrene &3$\times$3 &29.5 & 2 \\
    \bottomrule
    \end{tabular*}
\end{table*}

\subsection{Energy resolution}
\begin{figure}[!htb]
    \centering
    \includegraphics[width = 8.5cm]{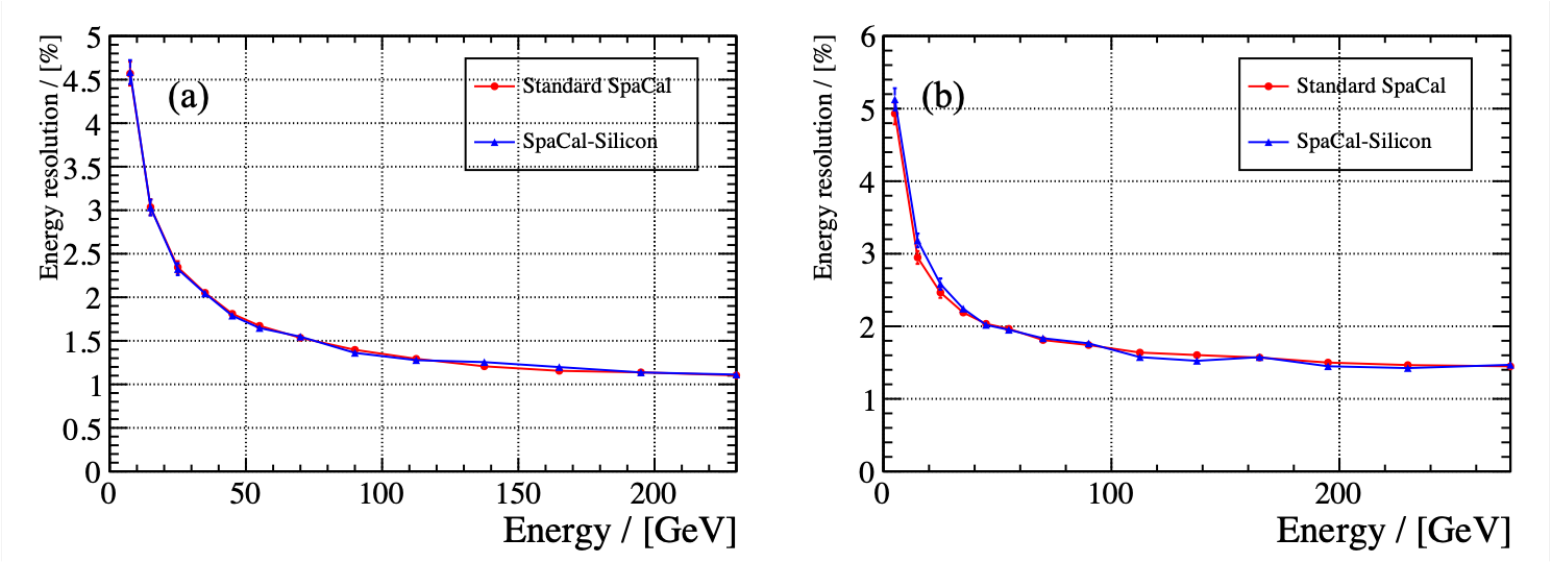}
    \caption{The energy resolution. (a): Pb-Polystyrene module vs. Pb-Polystyrene-Si module. (b): W-GAGG module vs. W-GAGG-Si module.}
    \label{fig:SPACAL_Silicon_ERes}
\end{figure}

For physics applications, the ability to preserve energy resolution is crucial because it directly impacts invariant mass reconstruction in channels such as $\pi^{0} \rightarrow \gamma \gamma$ or $\eta \rightarrow \gamma \gamma$. The hybrid design ensures that while spatial and temporal resolutions are improved, the essential calorimetric role of precise energy measurement is not compromised.
While the energy resolution degrades due to the inclusion of inactive material (primarily cooling layers) and the shortened upstream SpaCal, the addition of new active elements provides complementary information. Under these countervailing effects, the SpaCal-Silicon module achieves comparable energy resolution to the standard SpaCal module. Fig.~\ref{fig:SPACAL_Silicon_ERes} compares the energy resolution between two SpaCal-Silicon modules and standard SpaCal modules. It can be seen that the energy resolution does not deteriorate significantly, demonstrating that the SpaCal-Silicon design meets our preliminary resolution requirements.

\subsection{Position resolution}
\begin{figure}[!htb]
    \centering
    \includegraphics[width = 8.5cm]{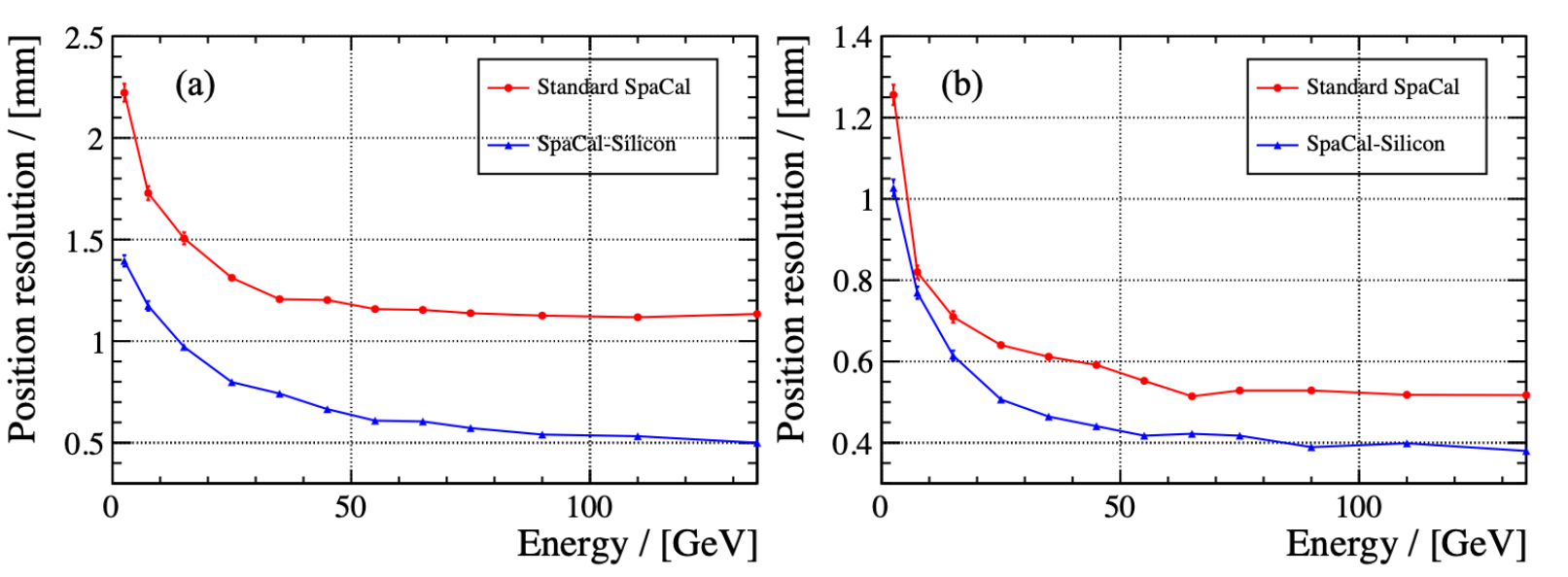}
    \caption{The position resolution. (a): Pb-Polystyrene module vs. Pb-Polystyrene-Si module. (b): W-GAGG module vs. W-GAGG-Si module.}
    \label{fig:SPACAL_Silicon_Pos_Res}
\end{figure}

The silicon pixel layers provide superior granularity compared to standard SpaCal modules, and granularity is strongly correlated with position resolution. We therefore expect improved position resolution in SpaCal-Silicon modules. Fig.~\ref{fig:SPACAL_Silicon_Pos_Res} compares the position resolution between two SpaCal-Silicon modules and standard SpaCal modules. As anticipated, the integration of silicon pixel layers significantly enhances the module's position resolution. The improvement becomes increasingly pronounced at higher photon energies, where electromagnetic showers are more compact and better contained. Under these conditions, the fine segmentation of silicon pixels significantly reduces the uncertainty in reconstructing the shower barycenter. This effect is particularly valuable for distinguishing nearby photon clusters originating from decays such as  $\pi^{0} \rightarrow \gamma \gamma$.

\subsection{Time information and resolution}

\begin{figure}[!htb]
    \centering
    \includegraphics[width = 8.5cm]{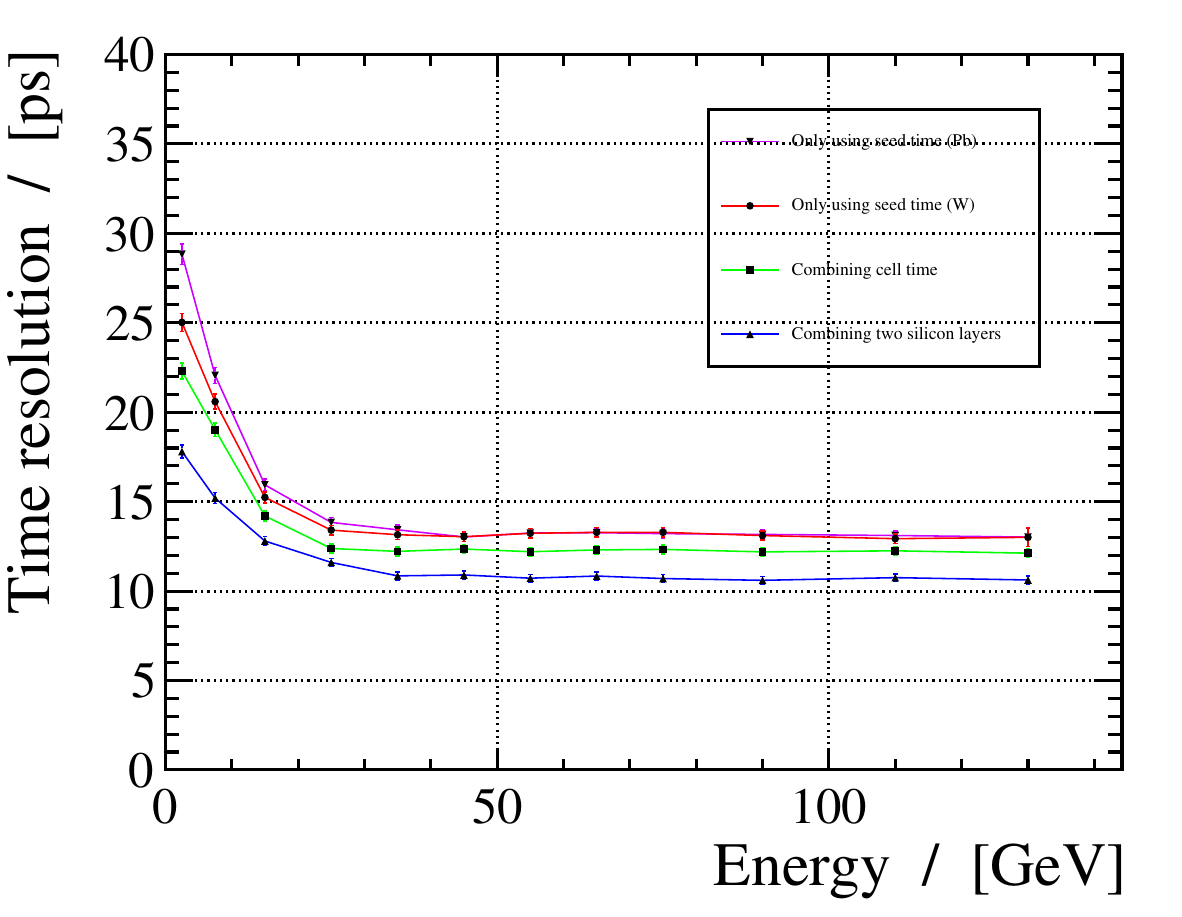}
    \caption{The purple and red lines show the time resolution of a single silicon layer using only seed timing information, with "W" and "Pb" representing modules W-GAGG-Si and Pb-Polystyrene-Si, respectively. The green line shows the time resolution after combining timing from multiple cells. The blue line represents the combined time resolution from two silicon layers.}
    \label{fig:硅像素层时间分辨}
\end{figure}

High-precision timing information is a critical parameter in high-luminosity collider physics experiments.Time resolution is not only a detector performance metric but also a physics enabler. In the HL-LHC environments, where pile-up leads to multiple simultaneous collisions per bunch crossing, the ability to assign precise time stamps to photons is essential for rejecting out-of-time background and improving vertex association. This section analyzes the timing characteristics of silicon pixel layers and their impact on overall module time resolution. Positioning silicon layers near the shower maximum ensures sufficient MIP flux at these locations, enabling high signal-to-noise ratio signals for improved timing precision. Fig.~\ref{fig:硅像素层时间分辨} shows the time resolution of silicon layers in different module configurations. In the low-energy region, silicon layers in W-GAGG-Si module exhibit superior time resolution compared to those in Pb-Polystyrene-Si module. The W-GAGG-Si module features a smaller Molière radius (dominated by the SpaCal parts in the module), resulting in higher MIP density per cell area compared to the Pb-polystyrene-Si module. This increased MIP flux through individual silicon pixels ultimately enhances timing resolution.

Based on previous analysis, a key consideration for silicon pixel layers with cell sizes smaller than the Molière radius is whether to combine timing information from multiple readout cells ($\mathrm{Cell^{2D}}$). This approach does not rely solely on the seed time of the cluster and has the potential to further improve the time resolution of the silicon pixel layer. In electromagnetic showers, transverse shower development occurs perpendicular to the direction of particle momentum. Consequently, the first-hit cell($\mathrm{Cell^{2D}}$), which typically has maximum energy, serves as the seed ($\mathrm{Seed^{2D}}$). As formalized in Equation~\ref{eq:Cell时间刻度}, a basic model assigns the timestamp of neighboring $\mathrm{Cell^{2D}}$ units as the $\mathrm{Seed^{2D}}$ timestamp plus a drift time determined by the distance between $\mathrm{Cell^{2D}}$ and $\mathrm{Seed^{2D}}$.

\begin{equation}\label{eq:Cell时间刻度}
    \begin{aligned}
        t_\mathrm{Cell}=t_\mathrm{Seed}+t_{Drift},
    \end{aligned}
\end{equation}

The extrapolated $\mathrm{Cell^{2D}}$ timestamp at $\mathrm{Seed^{2D}}$ is denoted as $t_{Cell}'$. The merged time information is then calculated via:

\begin{equation}\label{eq:合并时间信息}
    \begin{aligned}
        t=\frac{\sum_{i=0}^{n} t_{\mathrm{cell}_{i}}^{\prime} * E_{\mathrm{cell}_{i}}}{\sum_{i=0}^{n} E_{\mathrm{cell}_{i}}}.
    \end{aligned}
\end{equation}

Fig.~\ref{fig:硅像素层时间分辨} compares the time resolution between merged multi-$\mathrm{Cell^{2D}}$ timing and exclusive $\mathrm{Seed^{2D}}$ timing. It also presents the combined time resolution from all silicon layers, where the blue curve represents the SpaCal-Silicon module with two silicon layers.

\begin{figure}[!htb]
    \centering
    \includegraphics[width = 8.5cm]{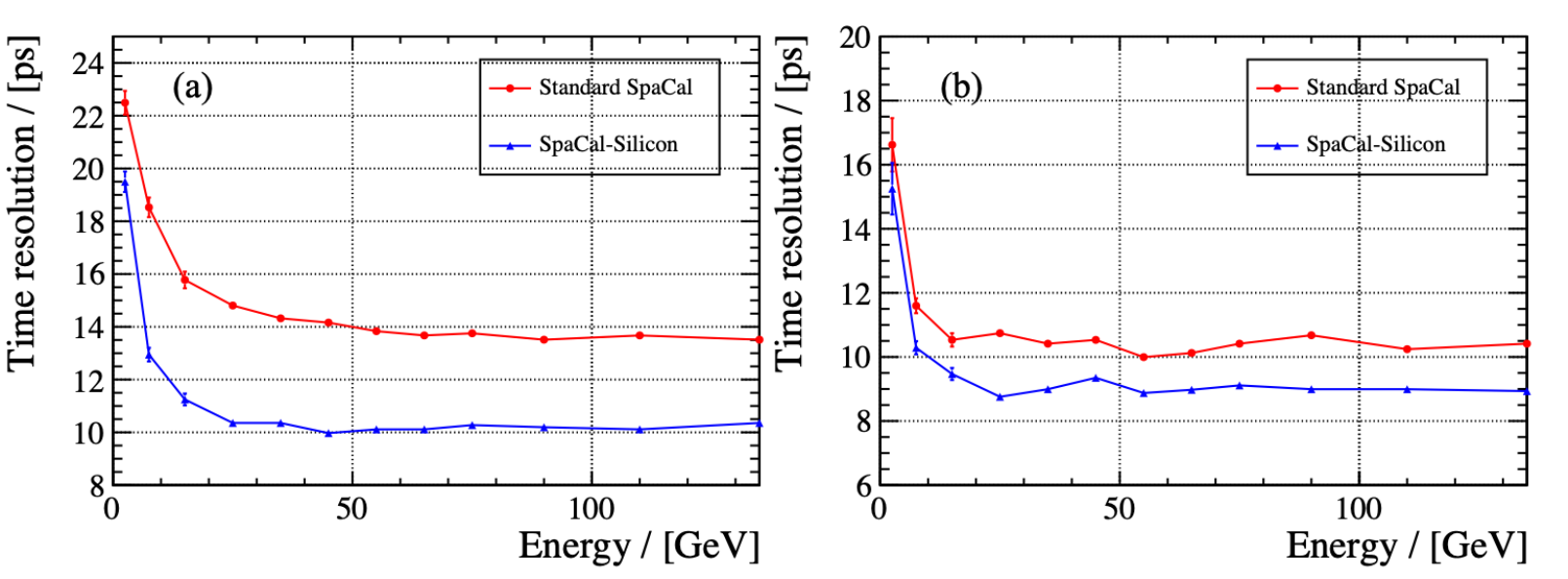}
    \caption{The time resolution. (a): Pb-Polystyrene module vs. Pb-Polystyrene-Si module. (b): W-GAGG module vs. W-GAGG-Si module.}
    \label{fig:SPACAL_Silicon_TRes}
\end{figure}

Fig.~\ref{fig:SPACAL_Silicon_TRes} presents the overall time resolution (merged timing information from both SpaCal and silicon layers) of two SpaCal-Silicon modules with comparison to the standard module. As shown in the plots, regardless of the type of SpaCal-Silicon module, it has achieved better time resolution compared to the standard SpaCal module.

\subsection{Performance of physical channels}
The primary purpose of electromagnetic calorimeters is to analyze final-state particles such as electrons, photons, $\pi^0$ mesons and their associated physical phenomena. It is crucial to emphasize that the resolution metrics only indirectly reflect detector performance in specific physical analyses. Systematic investigations of concrete physical processes therefore provide more direct insights into the operational efficacy of electromagnetic calorimeters. In this section, we will demonstrate the contribution of the SpaCal-Silicon module to the performance of specific physical channels through simulation experiments. For comparison, we also take the PicoCal in the LHCb Phase II upgrade as a concrete example and replace the standard SpaCal module in its central region with our newly designed SpaCal-Silicon module\cite{LHCbU2_FTDR,LHCbU2_ScopingDocu}. To evaluate the performance of the module under high-luminosity particle physics experiments, the simulation environment incorporates $pp$ collision backgrounds with instantaneous luminosity of $1.5\times 10^{34}\ \mathrm{cm^{-2}s^{-1}}$.

\subsubsection{Decay channels with high-energy photons in the final state}
\begin{figure}[!htb]
    \centering
    \includegraphics[width = 8.5cm]{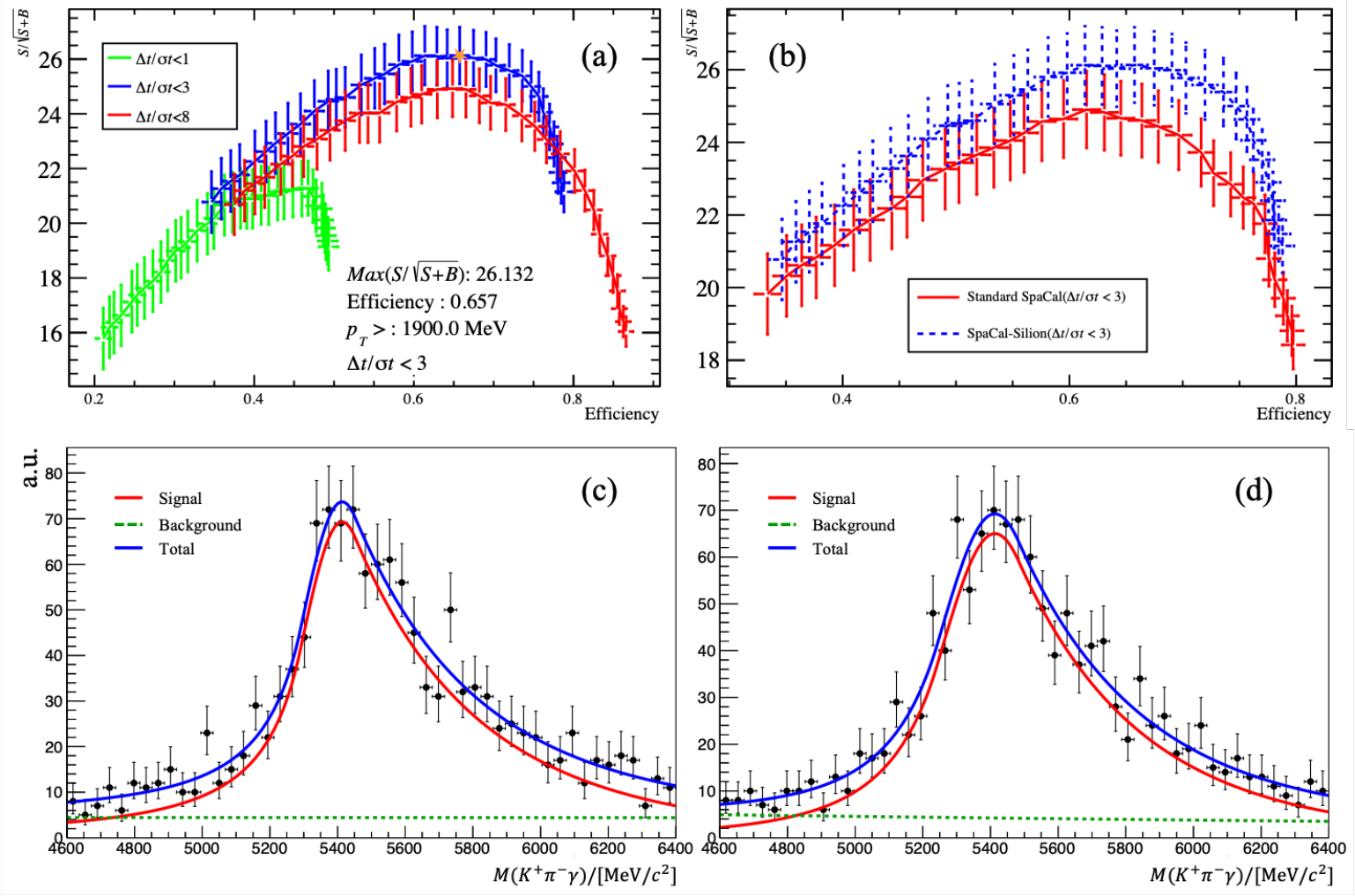}
    \caption{(a,b): The signal significance versus efficiency for $B^{0}\rightarrow K^{*0}\gamma$. (a): Signal significance through time window selection criteria in the SpaCal-Silicon module, with color-graded curves representing distinct time cut. The orange pentagram corresponding to the maximum signal significance. (b): Performance metrics between the standard SpaCal module and the SpaCal-Silicon module under a common dataset. Both configurations employ their respective optimized time window parameters that maximize signal significance. (c,d): Invariant mass distribution of $K^{*0}\gamma$ with photon reconstructed in the SpaCal region. (c): SpaCal-Silicon module. (d):Standrad SpaCal module.}
    \label{fig:B02KstGamma_Silicon}
\end{figure}

The $B^{0}\rightarrow K^{*0}\gamma$ decay channel is conventionally used in the LHCb experiment to validate calorimeter performance in high-energy photon detection. This channel features high-energy single photon emission, low background interference, reconstructible $K^{*0}\rightarrow K^+ \pi^-$ trajectories/vertices, and great Monte-Carlo data consistency, making it ideal for benchmarking. Fig.~\ref{fig:B02KstGamma_Silicon}a and b demonstrate the signal significance of $B^{0}\rightarrow K^{*0}\gamma$ within the SpaCal region, comparing the SpaCal-Silicon module configuration with the standard SpaCal module. The $K^{*0}$ candidates are reconstructed from combinations $K^+\pi^-$ with 10\% energy smearing applied to each track. The results indicate that, while energy resolution dominates the performance in this high-energy region, the fine time and position resolution of SpaCal-Silicon module still contributes marginally by improving vertex pointing of the photon. This effect helps to reduce systematic biases in invariant-mass reconstruction, which can otherwise smear the resonance peaks. Importantly, the performance gain, though modest, demonstrates that the hybrid design does not introduce a detrimental effect even in the region where it is not the primary driver of performance.

The texts in Fig.~\ref{fig:B02KstGamma_Silicon}a and b show the selection criteria, significance value, and efficiency when the signal significance is maximized, where $p_T$ represents the transverse momentum of the reconstructed photon, $\sigma t$ represents the time resolution of the module corresponding to the reconstructed photon energy. $\Delta t$ is defined as:
\begin{equation}\label{eq:时间Cut1}
    \begin{aligned}
        \Delta t = |t_{\gamma}'-t_{Gen_{K^{*0}}}|,
    \end{aligned}
\end{equation}

In the formula, $t_{Gen_{K^{*0}}}$ represents the time of the $K^{*0}$ production vertex, $t_{\gamma}'$ represents the photon time extrapolated to the $K^{*0}$ production vertex using time-of-flight information. The SpaCal-Silicon module demonstrates slightly improved performance in comparison to the standard SpaCal in hard-photon decay channels, as the performance in these channels is predominantly governed by energy resolution. Fig.~\ref{fig:B02KstGamma_Silicon}c and d show the invariant mass distribution of $K^{*0}\gamma$ corresponding to the maximum signal significance in Fig.~\ref{fig:B02KstGamma_Silicon}b.

\subsubsection{Decay channels with low-energy photons in the final state}
\begin{figure}[!htb]
    \centering
    \includegraphics[width = 8.5cm]{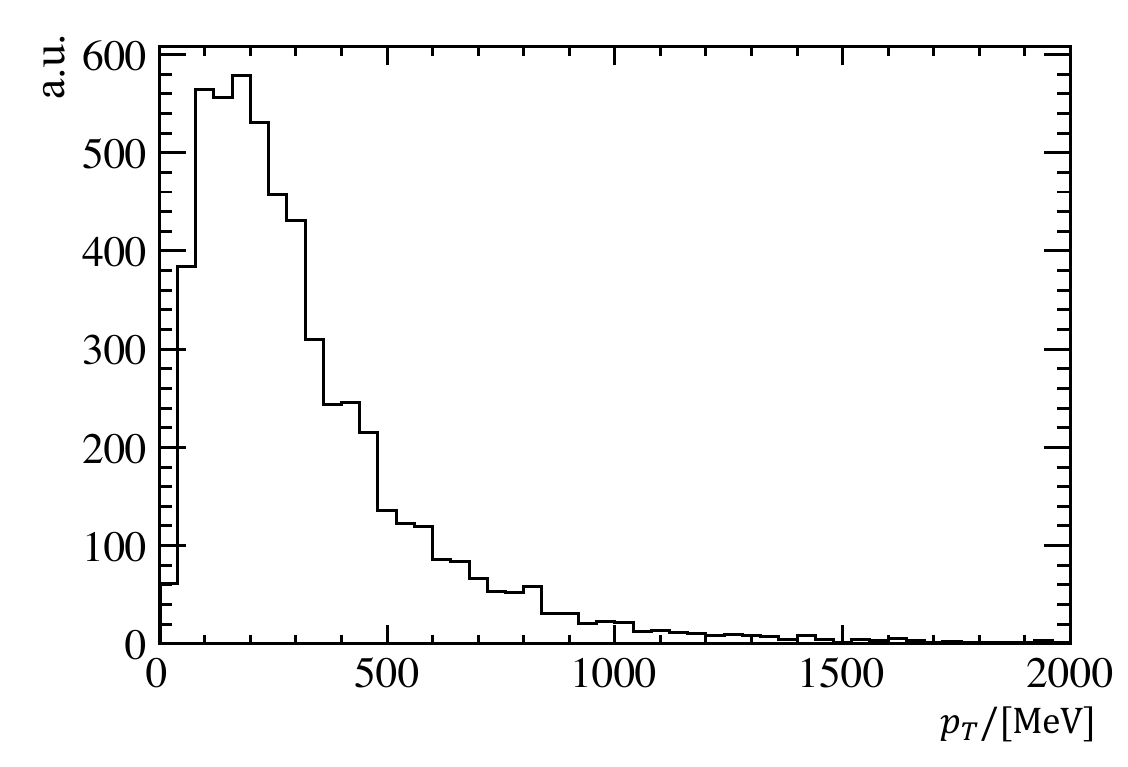}
    \caption{The transverse momentum of the photon from $B^-\rightarrow D^{*0}\pi^- \rightarrow (D^0,\gamma)\pi^-$ decay.}
    \label{fig:PtD0Gamma}
\end{figure}

$B^-\rightarrow D^{*0}\pi^- \rightarrow (D^0,\gamma)\pi^-$ decay is one of the key channels to probe the Standard Model and explore new physics. As shown in Fig.~\ref{fig:PtD0Gamma}, the photon from this channel has very low transverse momentum. Therefore, the precise time and position information provided by the SpaCal-Silicon module is expected to contribute to the signal selection and enhance the signal significance of this decay.

\begin{figure}[!htb]
    \centering
    \includegraphics[width = 8.5cm]{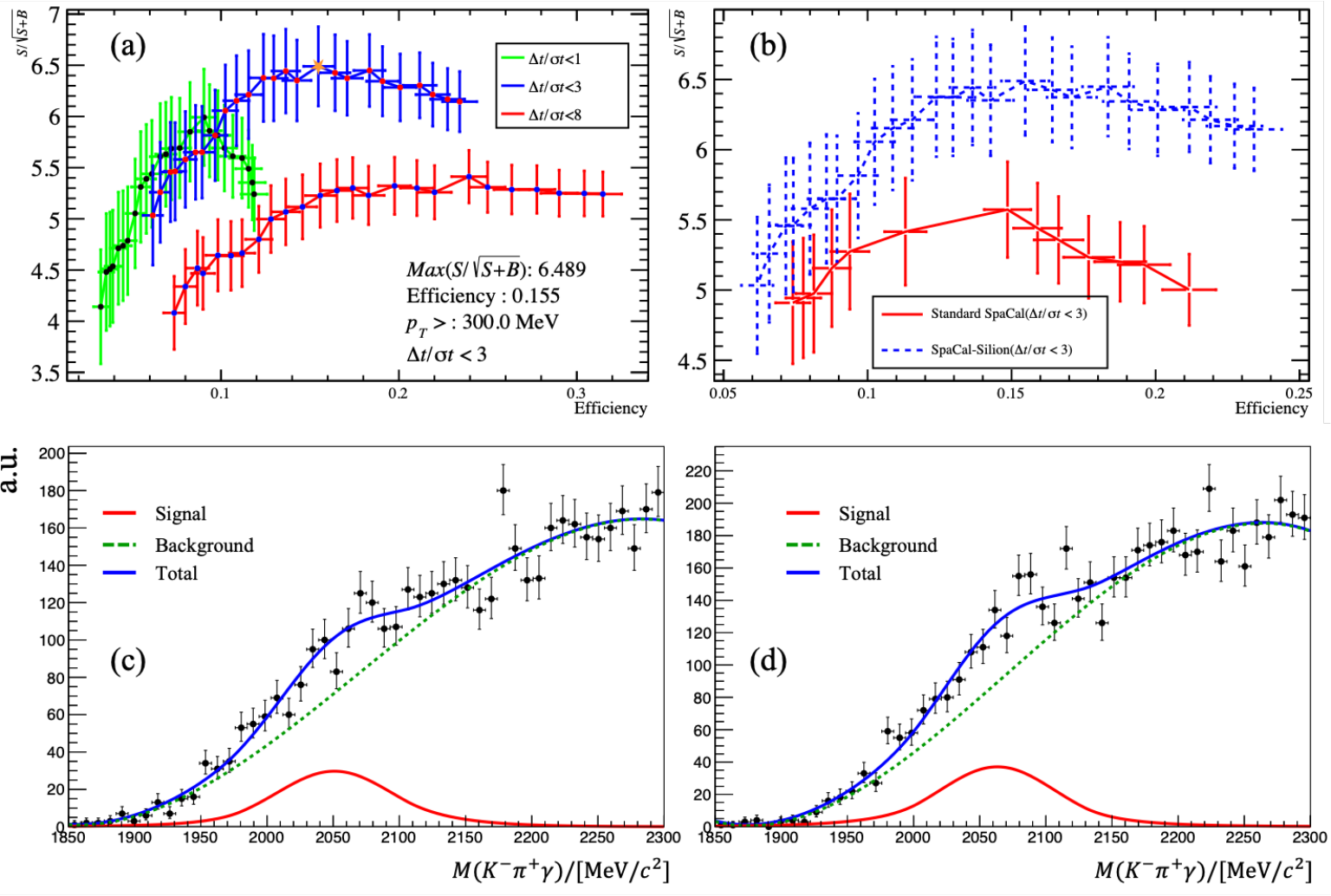}
    \caption{(a,b): The signal significance versus efficiency for $D^{*0}\rightarrow D^0 \gamma$. (a): Signal significance through time window selection criteria in the SpaCal-Silicon module, with color-graded curves representing distinct time cut. The orange pentagram corresponding to the maximum signal significance. (b): Performance metrics between the standard SpaCal module and the SpaCal-Silicon module under a common dataset. Both configurations employ their respective optimized time window parameters that maximize signal significance. (c,d): Invariant mass distribution of $D^0 \gamma$ with photon reconstructed in the SpaCal region. (c): SpaCal-Silicon module. (d):Standrad SpaCal module.}
    \label{fig:B2DstPi_Silicon}
\end{figure}

The $D^0$ candidates are reconstructed by combining $K^-\pi^+$ signal pairs, with 10\% Gaussian energy smearing applied to the $K^-/\pi^+$ mesons. Fig.~\ref{fig:B2DstPi_Silicon}a and b illustrate the relationship between signal significance and signal efficiency for the process $B^-\rightarrow D^{*0}(\rightarrow D^0 \gamma)\pi^-$ with $D^{*0}\rightarrow D^0 \gamma$. The selection criteria used in this analysis follow the definitions provided in the previous subsection.

Owing to superior time and position resolution, the SpaCal-Silicon module demonstrates enhanced performance in soft-photon decay channels. Fig.~\ref{fig:B2DstPi_Silicon}c and d show the invariant mass distribution of $D^0 \gamma$ corresponding to the maximum signal significance. This result highlights the physics relevance of the hybrid approach: while high-energy channels benefit only modestly, the low-energy photon regime critical for rare $B$ decays and $CP$ violation studies gains disproportionately from the improved granularity and timing. Such improvements could directly translate into the potential for new-physics discovery at both HL-LHC and next-generation $e^+e^-$ colliders.

\section{Conclusion}
In this paper, a hybrid design for integrating a silicon pixel layer into a SpaCal module is proposed for the performance enhancement needs of electromagnetic calorimeters in a high-luminosity collider experiment. 
The feasibility and performance benefits of the design are verified through a systematic simulation study. The design of the hybrid SpaCal-silicon module is based on embedding the silicon pixel layer into a longitudinally segmented SpaCal module, and the position of the silicon layer is optimized. Experimental validation shows that the best performance is achieved placing the silicon pixel layer at about 5-8 $X_0$ (radiation length); at the same time, the geometry of the silicon pixel layer is configured to be 5$\times$5 $\mathrm{mm^2}$ , in order to achieve higher granularity (smaller cell size) than the scintillator part. In addition, the two silicon layers share a single cooling layer, ensuring efficient energy deposition in the active material by minimizing material insertion.

To support the simulation study of this hybrid module, a simplified silicon pixel layer simulation model was developed based on the minimum number of MIP to establish a parametric relationship between energy deposition and voltage signal. A time noise model and a drift time correction method are also proposed to enable accurate simulation of the signal characteristics of the silicon pixel layer.

In terms of performance enhancement, the hybrid module significantly optimizes position resolution while maintaining energy resolution of the standard SpaCal module. For example, in the Pb-Polystyrene-Si configuration, the position resolution is improved by more than 56\%, while the time resolution is improved by about 14\% and 26\% in the W-GAGG-Si and Pb-Polystyrene-Si systems, respectively, which exhibit superior characteristics compared to the standard SpaCal module.

In the benchmarking of the decay channel, the hybrid module significantly improves the signal significance ($S/\sqrt{S+B}$) for $D^{*0}$ from the low-energy photon channel $B^-\rightarrow D^{*0}(\rightarrow D^0 \gamma)\pi^-$ by about 16\%; while in the high-energy photon channel, there is also a small performance improvement compared to the standard SpaCal module. In summary, the hybrid calorimeter module design embedded in the silicon pixel layer not only retains the advantages of the traditional scintillator module, but also significantly improves the multidimensional detection capability, provides a new technical solution for high-precision measurements in high-luminosity collider experiments, and provides a key reference for the development of future hybrid calorimeters.

\section{Acknowledgments}\label{sec.VII}
We express our gratitude to colleagues from the LHCb Upgrade R\&D group at CERN, Peking University and Tsinghua University for their helpful suggestions. In particular, Marco Pizzichemi, Loris Martinazzoli, and Philipp Gerhard Roloff developed the simulation software, and Zhenwei Yang, Liming Zhang and Zehua Xu provided insightful references that greatly inspired this study. Additionally, we extend our thanks to the developers and maintainers of the Gauss software in LHCb, which we used to generate events. The numerical calculations in this study were performed using a supercomputing system at the Supercomputing Center of Wuhan University.  This work is partially supported by National Natural Science Foundations of China under Grant No. W2443007.

\end{document}